\documentclass[aps,prl,twocolumn,amsmath,amssymb,superscriptaddress,floatfix,longbibliography]{revtex4-2}

\usepackage{graphicx}
\usepackage{multirow}
\usepackage[dvipsnames]{xcolor}
\usepackage{bm}
\usepackage{amsfonts}
\usepackage{dcolumn,braket,slashed}
\usepackage{times}
\usepackage{comment,ulem}

\begin{document}

\title{Tilt-Controlled Multi-Gap Superconductivity in Type-II Weyl Semimetals}

\author{Mingyu Lee}
\affiliation{Department of Physics, Hanyang University, Seoul 04763, South Korea}

\author{Myungjun Kang}
\email{kang43@uwm.edu}
\affiliation{Department of Physics, University of Wisconsin--Milwaukee, Milwaukee, Wisconsin 53201, USA}

\author{Sangmo Cheon}
\email{sangmocheon@hanyang.ac.kr}
\affiliation{Department of Physics, Hanyang University, Seoul 04763, South Korea}
\affiliation{High Pressure Research Center, Hanyang University, Seoul 04763, South Korea}
\affiliation{Research Institute for Natural Science, Hanyang University, Seoul 04763, South Korea}
\affiliation{Hanyang Institute for Quantum Science and Quantum Technology, Hanyang University, Seoul 04763, South Korea}

\begin{abstract}
In type-II Weyl semimetals, strongly tilted Weyl cones produce coexisting electron and hole pockets, a natural multi-band setting for superconductivity, yet how the tilt controls the number and relative size of the superconducting gaps remains unclear.
Here we classify the symmetry-allowed on-site pairing channels of a simple time-reversal symmetric tilted Weyl-semimetal model with $D_4$ symmetry and identify the fully symmetric $A_1$ channel as the leading instability for attractive on-site interactions, supporting a fully gapped two-gap state.
The tilt acts as a microscopic control parameter, redistributing the band-resolved pairing susceptibility between the two Fermi-surface sectors rather than simply rescaling the total density of states.
Once the tilt sufficiently enhances the secondary channel, a robust secondary gap develops and can exceed the primary gap even at weaker effective coupling, while inter-gap coupling locks both gaps to a common critical superconducting temperature.
The Weyl cone tilt thus controls the emergence and hierarchy of multi-gap superconductivity.
\end{abstract}

\maketitle

\textit{Introduction} --- 
Weyl semimetals (WSMs) provide a distinctive setting in which superconductivity develops on Fermi surfaces (FSs) endowed with chirality, Berry curvature, and topological surface connectivity~\cite{yan2017topological,armitage2018weyl,zhong2025weyl}.
Consequently, doping or external tuning can stabilize unconventional superconducting (SC) states whose gap structures are constrained jointly by crystalline symmetry and by the topology and geometry of the underlying Weyl bands~\cite{hosur2014time,wang2019superconductivity,bednik2015superconductivity}.
In type-I WSMs, this interplay has been studied in connection with nodal quasiparticle spectra, inter-node pairing, and finite-momentum pairing states such as Fulde--Ferrell--Larkin--Ovchinnikov or pair-density-wave phases~\cite{cho2012superconductivity,bednik2015superconductivity,li2018topological}.
These studies established that Weyl-node separation and FS topology can strongly restrict the possible SC states.

Type-II WSMs introduce a qualitatively different ingredient, a strong Weyl cone tilt.
Once the cones are overtilted, electron and hole pockets coexist at low energy and touch at the Weyl nodes~\cite{soluyanov2015type}.
The resulting pocketed electronic structure provides a natural multi-band platform in which distinct FS sectors may support distinct pairing amplitudes~\cite{suhl1959bardeen,moskalenko1959superconductivity}.
Theoretical studies of tilted Weyl superconductors have addressed nodal transitions, Majorana surface states, and enhanced SC responses near the type-I/type-II transition~\cite{alidoust2017superconductivity,li2017effect}.
However, the role of the tilt as a microscopic control parameter for multi-gap superconductivity, particularly how the tilt affects the pairing susceptibility, pairing channels, and the multiple SC gaps has not been established.

This question is motivated by recent experimental progress in Weyl-semimetal-based superconductors.
The noncentrosymmetric $T_d$ phases of MoTe$_2$ and WTe$_2$ have been widely discussed as candidate type-II WSMs~\cite{jiang2017signature,xu2018evidence,singh2020engineering,lee2026symmetry}.
Under external tuning, these and related type-II Weyl systems exhibit SC responses with multiple gap scales~\cite{naidyuk2018surface,mu2021pressure,zhang2024multigap} and, in some cases, evidence for multi-component pairing~\cite{guguchia2017signatures,luo2020possible}.
More broadly, recent experimental results have reported SC Fermi-arc surface states and distinct bulk and surface gap scales in WSM systems~\cite{kuibarov2024evidence,zhang2025atomic,schimmel2024surface}.

Here we develop such a framework using a simple time-reversal (TR) symmetric tilted WSM model based on a minimal Dirac semimetal (DSM) model.
This symmetry setting is particularly relevant to nonmagnetic, noncentrosymmetric type-II WSM candidates.
Time reversal (TR) maps a Weyl node at $\mathbf{k}$ to one at $-\mathbf{k}$ of the same chirality, so the vanishing of the total chirality requires at least four nodes~\cite{belopolski2017signatures,armitage2018weyl}, in contrast to the two-node structure that inversion symmetry permits in TR breaking models~\cite{mccormick2017minimal}.
We identify the fully symmetric $A_1$ representation as the leading on-site pairing channel, supporting a fully gapped multi-gap state within the parameter regime considered, and derive effective coupled gap equations for the resulting order parameters.
We find that the Weyl cone tilt does more than enhance the total low-energy density of states.
It redistributes the band-resolved pairing susceptibility and selectively strengthens a secondary pairing channel associated with the reconstructed FS, finally resulting in a reversal of the magnitude of the pairing channels.
This tilt-driven reversal of the SC gap hierarchy provides a microscopic and symmetry-based mechanism for controlling multi-gap superconductivity in Weyl semimetals.

\textit{Model Hamiltonian and Symmetry} ---
We first consider a simple type-II WSM, which is given by the Bloch Hamiltonian
\begin{eqnarray}
H_0(\mathbf{k})=t\cos k_z&+&t_0\left(\sin k_x\,\nu_x s_x+\sin k_y\,\nu_x s_y\right.\nonumber\\
&&\left.+\cos k_z\,\nu_z\right)+\lambda\sin k_z\,\nu_z s_z,
\label{eq:w20}
\end{eqnarray}
where $\nu_i$ and $s_i$ are Pauli matrices acting on the sublattice and
spin degrees of freedom, respectively.
This system arises from a minimal DSM where the Dirac cone is split due to a spin-orbit coupling like $\lambda$ term breaking the inversion symmetry, and then the Weyl cones are tilted due to the tilting $t$ term~\cite{soluyanov2015type,murakami2007phase}.

For simplicity, we work in the basis that diagonalizes the Dirac sector.
The pairing interaction itself remains on-site; the point-group classification is identical in the microscopic orbital basis~\cite{supple}.
In this basis, the normal-state
Hamiltonian can be written as
\begin{eqnarray}
H_N(\mathbf{k})=t\cos k_z+t_0 R(\mathbf{k})\tau_z+\lambda\sin k_z\tau_z\sigma_z ,
\label{eq:w2}
\end{eqnarray}
where $R(\mathbf{k})=\sqrt{\sin^2 k_x+\sin^2 k_y+\cos^2 k_z}$.
Here the Pauli matrices $\tau_i$ act on the two branches inherited from the parent DSM, and
$\sigma_i$ act on the TR related Kramers partners split by
inversion breaking.
The signs $\tau_z=\pm1$ label the two normal-state sectors that later support the primary and secondary SC gaps, respectively.

The SC phase is described by the BdG Hamiltonian
\begin{eqnarray}
\mathcal{H}_{\mathrm{BdG}}(\mathbf{k})
=
\begin{pmatrix}
H_N(\mathbf{k})-\mu & \Delta(\mathbf{k})\\
\Delta^\dagger(\mathbf{k}) & -H_N^T(-\mathbf{k})+\mu
\end{pmatrix},
\label{eq:BdG}
\end{eqnarray}
where $\mu$ is the chemical potential and $\Delta(\mathbf{k})$ is the SC pairing matrix.
The BdG Hamiltonian has the intrinsic particle-hole symmetry $\mathcal{C}=\eta_x K$, where $\eta_i$ acts in Nambu space.
For the TR invariant pairing channels considered below, the BdG Hamiltonian also preserves TR symmetry, and therefore belongs to symmetry class DIII~\cite{schnyder2008classification,chiu2016classification}.
These symmetry constraints determine the on-site pairing matrices and their classification under $D_4$.

\begin{table}[b]
\begin{tabular}{c c c c c c c}
 \hline
 ~~~~D$_4$~~~~& ~~~E~~~ & ~~2C$_4$~~ & ~~C$_2$~~ & ~~2C$'_2$~~ & ~~2C$''_2$~~ & ~~~~~~~~~~Pairing~~~~~~~~~~\\
 \hline
A$_1$ & 1 & 1 & 1 & 1 & 1 & $\sigma_y,\tau_x\sigma_y,\tau_z\sigma_y$\\
A$_2$ & 1 & 1 & 1 & -1 & -1 & $\tau_y\sigma_x$\\
B$_1$ & 1 & -1 & 1 & 1 & -1 & N/A\\
B$_2$ & 1 & -1 & 1 & -1 & 1 & N/A\\
E & 2 & 0 & -2 & 0 & 0 & $\tau_y,\tau_y\sigma_z$\\
 \hline
\end{tabular}
\caption{
\textbf{Character table of $D_4$ and classification of on-site pairing matrices.}
The first five columns give the characters of the single-valued irreducible representations of $D_4$.
The final column lists the momentum-independent pairing matrices transforming according to each representation.
N/A indicates that no on-site antisymmetric pairing matrix belongs to the corresponding representation.
}
\label{table1}
\end{table}

\textit{Classification of Pairing Channels} ---
We now classify the on-site (intra-unit-cell) SC pairing
matrices allowed by fermionic antisymmetry and the $D_4$ point-group
symmetry~\cite{sigrist1991phenomenological}.
We restrict the analysis to momentum-independent pairings, which provide
the minimal setting for studying multi-gap superconductivity in the
present model.
Experimentally, superconductivity in candidate Weyl-semimetal systems has
often been discussed in terms of $s$-wave or $s^\pm$-wave gap structures~\cite{guguchia2017signatures,naidyuk2018surface,lee2018origin}.
It should be noted that the transformation that transforms the Hamiltonian of Eq.~\eqref{eq:w20} to Eq.~\eqref{eq:w2} will result in momentum dependent terms.
We approximate these terms in each sector by its FS average, which results in constants~\cite{supple}.

The classification of the on-site SC pairings are summarized in Table~\ref{table1}, while the explicit transformation properties and the sector-resolved fermion bilinears are given in Table~S1~\cite{supple}.
Among the momentum-independent pairings considered here, only the
$A_1$ and $E$ representations contain more than one independent basis
matrix and can therefore support multiple gap amplitudes.
In the band basis, the $A_1$ representation is the only one that contains pairing matrices diagonal in the sector index, whereas the $E$ doublet connects the two sectors.

\begin{figure}[b]
\includegraphics[width=0.49\textwidth]{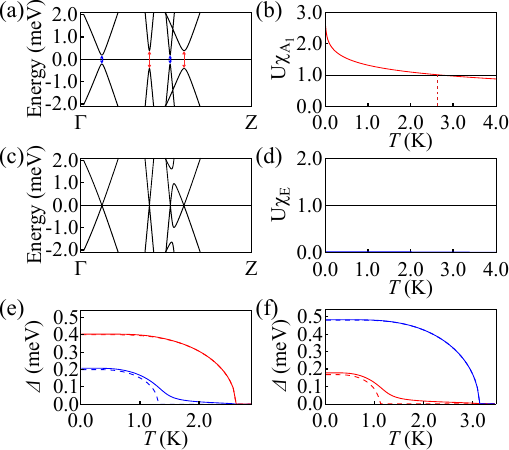}
\caption{\label{fig1:Pairing}
\textbf{Pairing channels in the simple model.}
\textbf{(a)} Band distribution and SC gap along the $\Gamma Z$ line for the $A_1$ channel at $T=0$.
The red and blue gaps denote $\Delta_1$ and $\Delta_2$, respectively,
corresponding to the $\tau=+1$ and $\tau=-1$ sectors.
The multi-gap SC state results in two gaps, induced via $\Delta_1$ and $\Delta_2$.
\textbf{(b)} Evolution of $U\chi_{A_1}$ with temperature for the $A_1$ channel.
The pairing instability is reached at the critical temperature marked by the vertical dotted line. 
\textbf{(c)} Band distribution along the $\Gamma Z$ line for the $E$ channel.
\textbf{(d)} Evolution of $U\chi_E$ with temperature for the $E$ channel. 
The pairing instability is never reached, therefore $T_c$ does not exist.
\textbf{(e)} Temperature dependence of $\Delta_1$ (red) and $\Delta_2$ (blue) for the $A_1$ channel when $t/t_0=1.5$, with (solid) and without (dotted) the inter-gap coupling.
The effective intra-band interactions are $U=4.93$~meV and $V=4.43$~meV, while the inter-gap coupling is $W=0.025$~meV.
\textbf{(f)} Temperature dependence of $\Delta_1$ (red) and $\Delta_2$ (blue) for the $A_1$ channel when $t/t_0=1.7$.
The parameters are $\mu=3$~meV, $t_0=10$~meV, and $\lambda=5$~meV.
}
\end{figure}

The first difference between the two channels is spectral.
Figures~\ref{fig1:Pairing}(a) and \ref{fig1:Pairing}(c) compare the $A_1$ and $E$ quasiparticle spectra at comparable gap amplitudes.
The $A_1$ channel produces a fully gapped multi-gap spectrum, whereas the $E$ channel, which pairs no time-reversed partners within a sector, remains gapless.

The second is the linearized instability.
As shown in Fig.~\ref{fig1:Pairing}(b), the leading eigenvalue of the $A_1$ pairing kernel reaches unity at a finite critical temperature $T_c$, whereas the corresponding $E$-channel eigenvalue stays below $2\times10^{-3}$ at every temperature shown in Fig.~\ref{fig1:Pairing}(d).
Held constant in the band basis, the $E$ vertex connects the two sectors exclusively and therefore acquires no FS pairing logarithm at all, which is why the separation from $A_1$ is so large.
We therefore identify the $A_1$ representation as the leading on-site multi-gap instability of the simple model of Eq.~\eqref{eq:BdG}; this conclusion remains unchanged when all on-site channels are compared under a common vertex normalization~\cite{supple}.

\textit{Coupled gap equations} --- 
Based on the classification Table, the most general on-site $A_1$ pairing matrix is
\begin{eqnarray}
\Delta(\mathbf{k})=
-i\left[\frac{\Delta_1+\Delta_2}{2}\sigma_y+\frac{\Delta_1-\Delta_2}{2}\tau_z\sigma_y+\Delta_3\tau_x\sigma_y\right],
\label{eq:A1_pairing}
\end{eqnarray}
where $\Delta_1$ and $\Delta_2$ describe intra-band pairing in the $\tau_z=1$ and $\tau_z=-1$ sectors, respectively, and directly gaps the FS.
The term $\Delta_3$ describes inter-band pairing between them, which does not gap the FS and is subdominant in the weak coupling regime.

Therefore, within the linearized three-component gap equation, $\Delta_3$ can be eliminated algebraically and absorbed into renormalized effective couplings between the two dominant intra-band order parameters~\cite{supple}.
The resulting two-gap equation takes the Suhl--Matthias--Walker form~\cite{suhl1959bardeen,moskalenko1959superconductivity},
\begin{eqnarray}
\begin{pmatrix}
\Delta_1\\
\Delta_2
\end{pmatrix}
=
\begin{pmatrix}
U\chi_1 & W\chi_2\\
W\chi_1 & V\chi_2
\end{pmatrix}
\begin{pmatrix}
\Delta_1\\
\Delta_2
\end{pmatrix},
\label{eq:effmulti}
\end{eqnarray}
where $U$ and $V$ are the effective attractive interactions within the two sectors, $W$ is the inter-gap coupling, and $\chi_i$ is the pairing susceptibility of sector $i$.
A SC instability occurs when the largest eigenvalue of the kernel in Eq.~\eqref{eq:effmulti} reaches unity, which is $U\chi_{A_1}$ of Fig.~\ref{fig1:Pairing}(b).

We solve the full nonlinear mean-field equations below $T_c$ and examine the pairing gaps $\Delta_1$, $\Delta_2$~\cite{supple}.
As shown in Fig.~\ref{fig1:Pairing}(e), the two pairings have distinct intrinsic transition temperatures for $W=0$, whereas finite $W$ locks both to a common $T_c$, producing the characteristic two-gap temperature dependence.
The BCS ratio for the two gaps are $2\Delta_1(0)/k_BT_c\simeq3.56$ and $2\Delta_2(0)/k_BT_c\simeq1.83$, compared with the single-band weak-coupling BCS value $3.53$.
Their opposite deviations are characteristic of coupled multi-gap SC systems~\cite{szabo2001evidence,bouquet2001specific,huang2007experimental,boaknin2003heat,kavcmarvcik2010studies,naidyuk2018surface,zhang2024multigap,luo2020possible}.

\textit{Tilt-induced Evolution of Multi-gap Superconductivity} ---
We next examine how the Weyl-cone tilt modifies the SC gap structure.
Although the normal-state type-I/type-II boundary occurs at $t/t_0=1$, the onset of a robust secondary gap is set by the pairing instability and need not coincide with this geometric boundary.
We characterize the spectrum by the minimum BdG excitation energy $E_{\min}=\min_{m,\mathbf{k}}|E_{m\mathbf{k}}|$, using $E_\text{th}=10~\mu$eV to distinguish weakly and robustly gapped regimes~\cite{supple}.

\begin{figure}[t]
\includegraphics[width=0.49\textwidth]{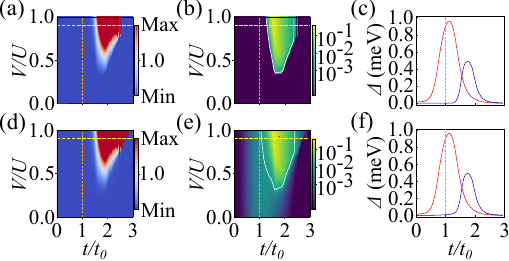}
\caption{\label{fig2:Phase}
\textbf{Tilt-driven evolution of the SC gap structure.}
\textbf{(a)} Ratio $\Delta_2/\Delta_1$ as a function of $V/U$ and Weyl-cone tilt $t/t_0$ for $W=0$.
Blue and red denote $\Delta_2<\Delta_1$ and $\Delta_2>\Delta_1$, respectively, with the white contour marking $\Delta_2=\Delta_1$.
\textbf{(b)} Corresponding minimum BdG gap $E_{\min}$, the white contour marks $E_\text{th}=10~\mu$eV, separating weakly gapped or numerically nodal spectra from the robustly gapped regime.
\textbf{(c)} $\Delta_1$ (red) and $\Delta_2$ (blue) versus $t/t_0$ along $V/U=0.9$.
\textbf{(d)-(f)} Same as \textbf{(a)-(c)} for $W=0.025$~meV.
Orange and yellow dashed lines mark $t/t_0=1$ and $V/U=0.9$, respectively.
All panels use $U=4.93$~meV and $T=0.05$~K.
The maps show pocketwise-constant gaps of the reduced model.
All other parameters are the same as in Fig.~\ref{fig1:Pairing}.
}
\end{figure}

Figures~\ref{fig2:Phase}(a-c) show the decoupled limit $W=0$.
At small tilt, the secondary susceptibility is weak and the corresponding gap remains below $E_\text{th}$.
Increasing tilt enhances this channel and drives a robust multi-gap state beyond an interaction-dependent threshold~[Fig.~\ref{fig2:Phase}(b)].
The equal-gap contour $\Delta_1=\Delta_2$ in Fig.~\ref{fig2:Phase}(a) further reveals a reversal from $\Delta_2<\Delta_1$ to $\Delta_2>\Delta_1$, even for $V/U<1$, showing that the gap hierarchy is not fixed by interaction strengths alone.
Along $V/U=0.9$ [Fig.~\ref{fig2:Phase}(c)], $\Delta_1$ peaks near $t/t_0\simeq1.1$, while $\Delta_2$ turns on at larger tilt, peaks near $t/t_0\simeq1.75$, and eventually exceeds $\Delta_1$.

For finite $W$~[Figs.~\ref{fig2:Phase}(d-f)], coupling to the primary sector induces a nonzero $\Delta_2$ even before the secondary channel becomes independently unstable.
At small tilt this gap remains below $E_\text{th}$ for weak $W$, producing a weakly gapped rather than nodal regime. Increasing $W$ mainly lifts this small gap, while leaving the gap-hierarchy reversal nearly unchanged~\cite{supple}.

Thus, the type-I/type-II boundary alone does not determine the SC gap topology.
Rather, tilt redistributes the pairing susceptibility between sectors, driving both the emergence of a robust secondary gap and the reversal of the gap hierarchy, as illustrated by Figs.~\ref{fig1:Pairing}(e,f).

\textit{Microscopic Origin} ---
We finally identify the microscopic origin of the tilt-driven gap evolution in Fig.~\ref{fig2:Phase}.
The $A_1$ pairing structure associates $\Delta_1$ and $\Delta_2$ with the $\tau=+1$ and $\tau=-1$ sectors of the normal-state Hamiltonian, respectively, while $\Delta_3$ couples the two sectors.
Projecting the linearized Cooper susceptibility onto the corresponding intra-sector vertices gives
\begin{eqnarray}
\chi_i(T)=\frac{1}{N}\sum_{\mathbf{k}}g_i(\mathbf{k}),
\end{eqnarray}
where $g_i(\mathbf{k})$ is the Cooper kernel restricted to sector $i=1,2$ within the pairing cutoff.

At small tilt, the pairing weight is concentrated in the primary sector and $\chi_1>\chi_2$.
Overtilting the Weyl cones reconstructs the electron and hole pockets and transfers low-energy pairing weight toward the secondary sector, causing $\chi_2$ to rapidly increase and eventually exceed $\chi_1$~\cite{supple}.
Thus, the tilt reverses the sector-resolved pairing-susceptibility hierarchy, providing the microscopic origin of the enhancement of $\Delta_2$ and the subsequent gap-hierarchy reversal observed in Fig.~\ref{fig2:Phase}.
The precise gap crossing remains determined by the full interaction kernel through $U$, $V$, and $W$.
By contrast, the inter-sector susceptibility associated with $\Delta_3$ lacks the Fermi-surface pairing logarithm because of the finite pair-energy mismatch and remains subdominant, supporting the effective two-gap description~\cite{supple}.

This channel-selective enhancement reflects a redistribution rather than a uniform increase of low-energy states.
The susceptibility ratio $\chi_2/\chi_1$ closely follows the sector-resolved DOS ratio $N_2/N_1$ as the tilt transfers spectral weight between the two sectors~\cite{supple}.
The gap-hierarchy reversal is therefore controlled primarily by the relative redistribution of low-energy spectral weight and pairing susceptibility, rather than by an overall enhancement of the total DOS.
This susceptibility reversal persists under variations of model parameters and when the filling, rather than the chemical potential, is held fixed~\cite{supple}.

\begin{figure}[t]
\includegraphics[width=0.49\textwidth]{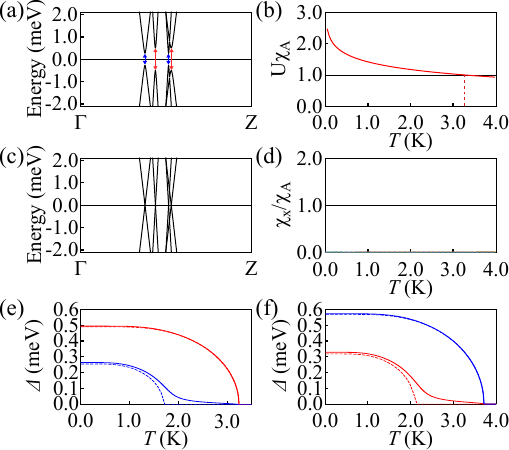}
\caption{\label{fig3:Minimal}
\textbf{Pairing channels in minimal models.}
\textbf{(a)} Band distribution and SC gap along the $\Gamma Z$ line for the $A$ channel at $T=0$.
The red and blue gaps denote $\Delta_1$ and $\Delta_2$, respectively,
corresponding to the $\tau=+1$ and $\tau=-1$ sectors.
The multi-gap SC state results in two gaps, induced via $\Delta_1$ and $\Delta_2$.
\textbf{(b)} Evolution of $U\chi_{A}$ with temperature for the $A$ channel.
The pairing instability is reached at the critical temperature marked by the vertical dotted line. 
\textbf{(c)} Band distribution along the $\Gamma Z$ line for the $B_2$ channel.
\textbf{(d)} Ratio $\chi_x/\chi_A$ for the competing channels $x$ =$B_1,B_2$ and $B_3$  at equal U. 
As there is no pairing instability, $T_c$ does not exist.
\textbf{(e)} Temperature dependence of $\Delta_1$ (red) and $\Delta_2$ (blue) for the $A$ channel when $t/t_0=2.95$, with (solid) and without (dotted) the inter-gap coupling.
The effective intra-band interactions are $U=16.3$~meV and $V=14.7$~meV, while the inter-gap coupling is $W=0.082$~meV.
\textbf{(f)} Temperature dependence of $\Delta_1$ (red) and $\Delta_2$ (blue) for the $A$ channel when $t/t_0=3.25$.
Details such as specific model parameters are presented in the Supplemental Material.
}
\end{figure}

\textit{Minimal Model} ---
To verify that the results above are not specific to the simple $D_4$ model, we finally consider a minimal TR-symmetric WSM with four Weyl nodes~\cite{supple}.
Unlike the 16-node simple model, given in Eq.~\eqref{eq:w20} the minimal model should have $D_2$ point-group symmetry.
Despite the different symmetry and Weyl-node structure, the fully symmetric $A$ representation remains the leading on-site multi-gap instability.
As shown in Fig.~\ref{fig3:Minimal}, the $A$ channel opens a full SC gap and reaches a finite-$T$ linearized instability, whereas the competing on-site channels remain subdominant.
Thus, the leading fully symmetric multi-gap instability found in the simple model persists in the minimal model.

\textit{Conclusion} ---
We have identified a symmetry-based mechanism for multi-gap superconductivity in a simple model of a TR symmetric tilted WSM.
The allowed on-site pairing channels select the $A_1$ representation, whose low-energy response is effectively described by coupled two-gap equations.
A finite inter-gap coupling locks the two gaps to a common $T_c$, while the Weyl-cone tilt redistributes the pairing susceptibility between the two FS sectors.
This redistribution can strongly enhance the secondary gap and even reverse the hierarchy, despite a weaker effective interaction.
Thus, unlike conventional two-band descriptions in which the gap hierarchy is governed primarily by interaction strengths~\cite{suhl1959bardeen,moskalenko1959superconductivity}, and earlier studies of the type-I/type-II transition with a single pairing channel~\cite{alidoust2017superconductivity,li2017effect}, here the normal-state band geometry controls the relative gap amplitudes.
The underlying susceptibility-redistribution mechanism is generic to multi-band systems near FS reconstruction~\cite{perali1996gap,innocenti2010resonant,coldea2018key}, while the Weyl structure fixes the relevant sectors, tilt, and pairing matrix elements.

This mechanism is relevant to SC type-II Weyl candidates such as $T_d$-MoTe$_2$ and WTe$_2$, whose electronic structures are tunable by pressure, strain, doping, and gating~\cite{soluyanov2015type,jiang2017signature,xu2018evidence,singh2020engineering}.
In MoTe$_2$, superconductivity is strongly enhanced under pressure and multicomponent SC responses have been reported~\cite{qi2016superconductivity,zhang2024multigap,luo2020possible,naidyuk2018surface,guguchia2017signatures}, while its normal-state band topology is well established~\cite{bahramy2018ubiquitous,wang2019higher,lee2026symmetry}.
The proposed mechanism can be tested by correlating the evolution of the FSs~\cite{jiang2017signature,xu2018evidence,hu2020detection} with spectroscopic or thermodynamic measurements of the SC gaps~\cite{naidyuk2018surface,luo2020possible,zhang2024multigap,huang2007experimental,bouquet2001specific}.
Material-specific interactions, disorder, surface channels~\cite{xu2014topological,kuibarov2024evidence,schimmel2024surface}, competing pairing states~\cite{lane2022identifying,schrade2024nematic}, and finite-momentum pairing~\cite{cho2012superconductivity} may modify quantitative phase boundaries, but not the central principle; normal-state band geometry can control both the emergence and the relative hierarchy of multiple SC gaps.
\\

\textit{Acknowledgements}---
This work was supported by the National Research Foundation of South Korea (NRF) through the Basic Science Research Programs (Grant Nos. NRF-2022R1A2C1011646, NRF-2022M3H3A1085772, RS-2024-00416036, RS-2025-03392969, and RS-2026-25490114), by the Creation of the Quantum Information Science R\&D Ecosystem through the NRF funded by the South Korean government (Ministry of Science and ICT) (Grant No. RS-2023-NR068116), by the Quantum Simulator Development Project for Materials Innovation through the NRF funded by the Ministry of Science and ICT, Republic of South Korea (Grant No. RS-2023-NR119931), by the Brain Pool Program funded by the Ministry of Science and ICT through the National Research Foundation of South Korea (Grant No. RS-2025-25446099), and by the Institute of Information \& Communications Technology Planning \& Evaluation (IITP) through the Information Technology Research Center (ITRC) program funded by the Ministry of Science and ICT, Republic of South Korea, under the project ``Quantum--AI Convergence for Innovation and Talent Development'' (Grant No. RS-2026-25519864).
M. Kang was supported by the National Science Foundation Grant No. DMREF 2323857.
\\

\textit{Data availability} ---
The data that support the findings of this study are available from the corresponding author upon reasonable request.


\begin{thebibliography}{48}%
\makeatletter
\providecommand \@ifxundefined [1]{%
 \@ifx{#1\undefined}
}%
\providecommand \@ifnum [1]{%
 \ifnum #1\expandafter \@firstoftwo
 \else \expandafter \@secondoftwo
 \fi
}%
\providecommand \@ifx [1]{%
 \ifx #1\expandafter \@firstoftwo
 \else \expandafter \@secondoftwo
 \fi
}%
\providecommand \natexlab [1]{#1}%
\providecommand \enquote  [1]{``#1''}%
\providecommand \bibnamefont  [1]{#1}%
\providecommand \bibfnamefont [1]{#1}%
\providecommand \citenamefont [1]{#1}%
\providecommand \href@noop [0]{\@secondoftwo}%
\providecommand \href [0]{\begingroup \@sanitize@url \@href}%
\providecommand \@href[1]{\@@startlink{#1}\@@href}%
\providecommand \@@href[1]{\endgroup#1\@@endlink}%
\providecommand \@sanitize@url [0]{\catcode `\\12\catcode `\$12\catcode `\&12\catcode `\#12\catcode `\^12\catcode `\_12\catcode `\%12\relax}%
\providecommand \@@startlink[1]{}%
\providecommand \@@endlink[0]{}%
\providecommand \url  [0]{\begingroup\@sanitize@url \@url }%
\providecommand \@url [1]{\endgroup\@href {#1}{\urlprefix }}%
\providecommand \urlprefix  [0]{URL }%
\providecommand \Eprint [0]{\href }%
\providecommand \doibase [0]{https://doi.org/}%
\providecommand \selectlanguage [0]{\@gobble}%
\providecommand \bibinfo  [0]{\@secondoftwo}%
\providecommand \bibfield  [0]{\@secondoftwo}%
\providecommand \translation [1]{[#1]}%
\providecommand \BibitemOpen [0]{}%
\providecommand \bibitemStop [0]{}%
\providecommand \bibitemNoStop [0]{.\EOS\space}%
\providecommand \EOS [0]{\spacefactor3000\relax}%
\providecommand \BibitemShut  [1]{\csname bibitem#1\endcsname}%
\let\auto@bib@innerbib\@empty
\bibitem [{\citenamefont {Yan}\ and\ \citenamefont {Felser}(2017)}]{yan2017topological}%
  \BibitemOpen
  \bibfield  {author} {\bibinfo {author} {\bibfnamefont {B.}~\bibnamefont {Yan}}\ and\ \bibinfo {author} {\bibfnamefont {C.}~\bibnamefont {Felser}},\ }\bibfield  {title} {\bibinfo {title} {{Topological materials: Weyl semimetals}},\ }\href@noop {} {\bibfield  {journal} {\bibinfo  {journal} {Annual Review of Condensed Matter Physics}\ }\textbf {\bibinfo {volume} {8}},\ \bibinfo {pages} {337} (\bibinfo {year} {2017})}\BibitemShut {NoStop}%
\bibitem [{\citenamefont {Armitage}\ \emph {et~al.}(2018)\citenamefont {Armitage}, \citenamefont {Mele},\ and\ \citenamefont {Vishwanath}}]{armitage2018weyl}%
  \BibitemOpen
  \bibfield  {author} {\bibinfo {author} {\bibfnamefont {N.~P.}\ \bibnamefont {Armitage}}, \bibinfo {author} {\bibfnamefont {E.~J.}\ \bibnamefont {Mele}},\ and\ \bibinfo {author} {\bibfnamefont {A.}~\bibnamefont {Vishwanath}},\ }\bibfield  {title} {\bibinfo {title} {{Weyl and Dirac semimetals in three-dimensional solids}},\ }\href@noop {} {\bibfield  {journal} {\bibinfo  {journal} {Reviews of Modern Physics}\ }\textbf {\bibinfo {volume} {90}},\ \bibinfo {pages} {015001} (\bibinfo {year} {2018})}\BibitemShut {NoStop}%
\bibitem [{\citenamefont {Zhong}\ \emph {et~al.}(2025)\citenamefont {Zhong}, \citenamefont {Vu}, \citenamefont {Zhai}, \citenamefont {Soh}, \citenamefont {Liu}, \citenamefont {Wu}, \citenamefont {Suwardi}, \citenamefont {Liu}, \citenamefont {Chang}, \citenamefont {Loh} \emph {et~al.}}]{zhong2025weyl}%
  \BibitemOpen
  \bibfield  {author} {\bibinfo {author} {\bibfnamefont {M.}~\bibnamefont {Zhong}}, \bibinfo {author} {\bibfnamefont {N.~T.~T.}\ \bibnamefont {Vu}}, \bibinfo {author} {\bibfnamefont {W.}~\bibnamefont {Zhai}}, \bibinfo {author} {\bibfnamefont {J.~R.}\ \bibnamefont {Soh}}, \bibinfo {author} {\bibfnamefont {Y.}~\bibnamefont {Liu}}, \bibinfo {author} {\bibfnamefont {J.}~\bibnamefont {Wu}}, \bibinfo {author} {\bibfnamefont {A.}~\bibnamefont {Suwardi}}, \bibinfo {author} {\bibfnamefont {H.}~\bibnamefont {Liu}}, \bibinfo {author} {\bibfnamefont {G.}~\bibnamefont {Chang}}, \bibinfo {author} {\bibfnamefont {K.~P.}\ \bibnamefont {Loh}}, \emph {et~al.},\ }\bibfield  {title} {\bibinfo {title} {{Weyl semimetals: From principles, materials to applications}},\ }\href@noop {} {\bibfield  {journal} {\bibinfo  {journal} {Advanced Materials}\ }\textbf {\bibinfo {volume} {37}},\ \bibinfo {pages} {2506236} (\bibinfo {year} {2025})}\BibitemShut {NoStop}%
\bibitem [{\citenamefont {Hosur}\ \emph {et~al.}(2014)\citenamefont {Hosur}, \citenamefont {Dai}, \citenamefont {Fang},\ and\ \citenamefont {Qi}}]{hosur2014time}%
  \BibitemOpen
  \bibfield  {author} {\bibinfo {author} {\bibfnamefont {P.}~\bibnamefont {Hosur}}, \bibinfo {author} {\bibfnamefont {X.}~\bibnamefont {Dai}}, \bibinfo {author} {\bibfnamefont {Z.}~\bibnamefont {Fang}},\ and\ \bibinfo {author} {\bibfnamefont {X.-L.}\ \bibnamefont {Qi}},\ }\bibfield  {title} {\bibinfo {title} {{Time-reversal-invariant topological superconductivity in doped Weyl semimetals}},\ }\href@noop {} {\bibfield  {journal} {\bibinfo  {journal} {Physical Review B}\ }\textbf {\bibinfo {volume} {90}},\ \bibinfo {pages} {045130} (\bibinfo {year} {2014})}\BibitemShut {NoStop}%
\bibitem [{\citenamefont {Wang}(2019)}]{wang2019superconductivity}%
  \BibitemOpen
  \bibfield  {author} {\bibinfo {author} {\bibfnamefont {J.}~\bibnamefont {Wang}},\ }\bibfield  {title} {\bibinfo {title} {{Superconductivity in topological semimetals}},\ }\href@noop {} {\bibfield  {journal} {\bibinfo  {journal} {National Science Review}\ }\textbf {\bibinfo {volume} {6}},\ \bibinfo {pages} {199} (\bibinfo {year} {2019})}\BibitemShut {NoStop}%
\bibitem [{\citenamefont {Bednik}\ \emph {et~al.}(2015)\citenamefont {Bednik}, \citenamefont {Zyuzin},\ and\ \citenamefont {Burkov}}]{bednik2015superconductivity}%
  \BibitemOpen
  \bibfield  {author} {\bibinfo {author} {\bibfnamefont {G.}~\bibnamefont {Bednik}}, \bibinfo {author} {\bibfnamefont {A.}~\bibnamefont {Zyuzin}},\ and\ \bibinfo {author} {\bibfnamefont {A.}~\bibnamefont {Burkov}},\ }\bibfield  {title} {\bibinfo {title} {{Superconductivity in Weyl metals}},\ }\href@noop {} {\bibfield  {journal} {\bibinfo  {journal} {Physical Review B}\ }\textbf {\bibinfo {volume} {92}},\ \bibinfo {pages} {035153} (\bibinfo {year} {2015})}\BibitemShut {NoStop}%
\bibitem [{\citenamefont {Cho}\ \emph {et~al.}(2012)\citenamefont {Cho}, \citenamefont {Bardarson}, \citenamefont {Lu},\ and\ \citenamefont {Moore}}]{cho2012superconductivity}%
  \BibitemOpen
  \bibfield  {author} {\bibinfo {author} {\bibfnamefont {G.~Y.}\ \bibnamefont {Cho}}, \bibinfo {author} {\bibfnamefont {J.~H.}\ \bibnamefont {Bardarson}}, \bibinfo {author} {\bibfnamefont {Y.-M.}\ \bibnamefont {Lu}},\ and\ \bibinfo {author} {\bibfnamefont {J.~E.}\ \bibnamefont {Moore}},\ }\bibfield  {title} {\bibinfo {title} {{Superconductivity of doped Weyl semimetals: Finite-momentum pairing and electronic analog of the \textsuperscript{3}He-A phase}},\ }\href@noop {} {\bibfield  {journal} {\bibinfo  {journal} {Physical Review B}\ }\textbf {\bibinfo {volume} {86}},\ \bibinfo {pages} {214514} (\bibinfo {year} {2012})}\BibitemShut {NoStop}%
\bibitem [{\citenamefont {Li}\ and\ \citenamefont {Haldane}(2018)}]{li2018topological}%
  \BibitemOpen
  \bibfield  {author} {\bibinfo {author} {\bibfnamefont {Y.}~\bibnamefont {Li}}\ and\ \bibinfo {author} {\bibfnamefont {F.~D.~M.}\ \bibnamefont {Haldane}},\ }\bibfield  {title} {\bibinfo {title} {{Topological Nodal Cooper Pairing in Doped Weyl Metals}},\ }\href@noop {} {\bibfield  {journal} {\bibinfo  {journal} {Physical Review Letters}\ }\textbf {\bibinfo {volume} {120}},\ \bibinfo {pages} {067003} (\bibinfo {year} {2018})}\BibitemShut {NoStop}%
\bibitem [{\citenamefont {Soluyanov}\ \emph {et~al.}(2015)\citenamefont {Soluyanov}, \citenamefont {Gresch}, \citenamefont {Wang}, \citenamefont {Wu}, \citenamefont {Troyer}, \citenamefont {Dai},\ and\ \citenamefont {Bernevig}}]{soluyanov2015type}%
  \BibitemOpen
  \bibfield  {author} {\bibinfo {author} {\bibfnamefont {A.~A.}\ \bibnamefont {Soluyanov}}, \bibinfo {author} {\bibfnamefont {D.}~\bibnamefont {Gresch}}, \bibinfo {author} {\bibfnamefont {Z.}~\bibnamefont {Wang}}, \bibinfo {author} {\bibfnamefont {Q.}~\bibnamefont {Wu}}, \bibinfo {author} {\bibfnamefont {M.}~\bibnamefont {Troyer}}, \bibinfo {author} {\bibfnamefont {X.}~\bibnamefont {Dai}},\ and\ \bibinfo {author} {\bibfnamefont {B.~A.}\ \bibnamefont {Bernevig}},\ }\bibfield  {title} {\bibinfo {title} {{Type-II Weyl Semimetals}},\ }\href@noop {} {\bibfield  {journal} {\bibinfo  {journal} {Nature}\ }\textbf {\bibinfo {volume} {527}},\ \bibinfo {pages} {495} (\bibinfo {year} {2015})}\BibitemShut {NoStop}%
\bibitem [{\citenamefont {Suhl}\ \emph {et~al.}(1959)\citenamefont {Suhl}, \citenamefont {Matthias},\ and\ \citenamefont {Walker}}]{suhl1959bardeen}%
  \BibitemOpen
  \bibfield  {author} {\bibinfo {author} {\bibfnamefont {H.}~\bibnamefont {Suhl}}, \bibinfo {author} {\bibfnamefont {B.}~\bibnamefont {Matthias}},\ and\ \bibinfo {author} {\bibfnamefont {L.}~\bibnamefont {Walker}},\ }\bibfield  {title} {\bibinfo {title} {{Bardeen-Cooper-Schrieffer theory of superconductivity in the case of overlapping bands}},\ }\href@noop {} {\bibfield  {journal} {\bibinfo  {journal} {Physical Review Letters}\ }\textbf {\bibinfo {volume} {3}},\ \bibinfo {pages} {552} (\bibinfo {year} {1959})}\BibitemShut {NoStop}%
\bibitem [{\citenamefont {Moskalenko}(1959)}]{moskalenko1959superconductivity}%
  \BibitemOpen
  \bibfield  {author} {\bibinfo {author} {\bibfnamefont {V.~A.}\ \bibnamefont {Moskalenko}},\ }\bibfield  {title} {\bibinfo {title} {{Superconductivity in metals with overlapping energy bands}},\ }\href@noop {} {\bibfield  {journal} {\bibinfo  {journal} {Fiz. Met. Metalloved.}\ }\textbf {\bibinfo {volume} {8}},\ \bibinfo {pages} {503} (\bibinfo {year} {1959})}\BibitemShut {NoStop}%
\bibitem [{\citenamefont {Alidoust}\ \emph {et~al.}(2017)\citenamefont {Alidoust}, \citenamefont {Halterman},\ and\ \citenamefont {Zyuzin}}]{alidoust2017superconductivity}%
  \BibitemOpen
  \bibfield  {author} {\bibinfo {author} {\bibfnamefont {M.}~\bibnamefont {Alidoust}}, \bibinfo {author} {\bibfnamefont {K.}~\bibnamefont {Halterman}},\ and\ \bibinfo {author} {\bibfnamefont {A.}~\bibnamefont {Zyuzin}},\ }\bibfield  {title} {\bibinfo {title} {{Superconductivity in type-II Weyl semimetals}},\ }\href@noop {} {\bibfield  {journal} {\bibinfo  {journal} {Physical Review B}\ }\textbf {\bibinfo {volume} {95}},\ \bibinfo {pages} {155124} (\bibinfo {year} {2017})}\BibitemShut {NoStop}%
\bibitem [{\citenamefont {Li}\ \emph {et~al.}(2017)\citenamefont {Li}, \citenamefont {Rosenstein}, \citenamefont {Shapiro},\ and\ \citenamefont {Shapiro}}]{li2017effect}%
  \BibitemOpen
  \bibfield  {author} {\bibinfo {author} {\bibfnamefont {D.}~\bibnamefont {Li}}, \bibinfo {author} {\bibfnamefont {B.}~\bibnamefont {Rosenstein}}, \bibinfo {author} {\bibfnamefont {B.~Y.}\ \bibnamefont {Shapiro}},\ and\ \bibinfo {author} {\bibfnamefont {I.}~\bibnamefont {Shapiro}},\ }\bibfield  {title} {\bibinfo {title} {{Effect of the type-I to type-II Weyl semimetal topological transition on superconductivity}},\ }\href@noop {} {\bibfield  {journal} {\bibinfo  {journal} {Physical Review B}\ }\textbf {\bibinfo {volume} {95}},\ \bibinfo {pages} {094513} (\bibinfo {year} {2017})}\BibitemShut {NoStop}%
\bibitem [{\citenamefont {Jiang}\ \emph {et~al.}(2017)\citenamefont {Jiang}, \citenamefont {Liu}, \citenamefont {Sun}, \citenamefont {Yang}, \citenamefont {Rajamathi}, \citenamefont {Qi}, \citenamefont {Yang}, \citenamefont {Chen}, \citenamefont {Peng}, \citenamefont {Hwang} \emph {et~al.}}]{jiang2017signature}%
  \BibitemOpen
  \bibfield  {author} {\bibinfo {author} {\bibfnamefont {J.}~\bibnamefont {Jiang}}, \bibinfo {author} {\bibfnamefont {Z.}~\bibnamefont {Liu}}, \bibinfo {author} {\bibfnamefont {Y.}~\bibnamefont {Sun}}, \bibinfo {author} {\bibfnamefont {H.}~\bibnamefont {Yang}}, \bibinfo {author} {\bibfnamefont {C.}~\bibnamefont {Rajamathi}}, \bibinfo {author} {\bibfnamefont {Y.}~\bibnamefont {Qi}}, \bibinfo {author} {\bibfnamefont {L.}~\bibnamefont {Yang}}, \bibinfo {author} {\bibfnamefont {C.}~\bibnamefont {Chen}}, \bibinfo {author} {\bibfnamefont {H.}~\bibnamefont {Peng}}, \bibinfo {author} {\bibfnamefont {C.}~\bibnamefont {Hwang}}, \emph {et~al.},\ }\bibfield  {title} {\bibinfo {title} {{Signature of type-II Weyl semimetal phase in MoTe\textsubscript{2}}},\ }\href@noop {} {\bibfield  {journal} {\bibinfo  {journal} {Nature Communications}\ }\textbf {\bibinfo {volume} {8}},\ \bibinfo {pages} {13973} (\bibinfo {year} {2017})}\BibitemShut {NoStop}%
\bibitem [{\citenamefont {Xu}\ \emph {et~al.}(2018)\citenamefont {Xu}, \citenamefont {Wang}, \citenamefont {Magrez}, \citenamefont {Bugnon}, \citenamefont {Berger}, \citenamefont {Matt}, \citenamefont {Strocov}, \citenamefont {Plumb}, \citenamefont {Radovic}, \citenamefont {Pomjakushina} \emph {et~al.}}]{xu2018evidence}%
  \BibitemOpen
  \bibfield  {author} {\bibinfo {author} {\bibfnamefont {N.}~\bibnamefont {Xu}}, \bibinfo {author} {\bibfnamefont {Z.}~\bibnamefont {Wang}}, \bibinfo {author} {\bibfnamefont {A.}~\bibnamefont {Magrez}}, \bibinfo {author} {\bibfnamefont {P.}~\bibnamefont {Bugnon}}, \bibinfo {author} {\bibfnamefont {H.}~\bibnamefont {Berger}}, \bibinfo {author} {\bibfnamefont {C.~E.}\ \bibnamefont {Matt}}, \bibinfo {author} {\bibfnamefont {V.~N.}\ \bibnamefont {Strocov}}, \bibinfo {author} {\bibfnamefont {N.~C.}\ \bibnamefont {Plumb}}, \bibinfo {author} {\bibfnamefont {M.}~\bibnamefont {Radovic}}, \bibinfo {author} {\bibfnamefont {E.}~\bibnamefont {Pomjakushina}}, \emph {et~al.},\ }\bibfield  {title} {\bibinfo {title} {{Evidence of a Coulomb-interaction-induced Lifshitz transition and robust hybrid Weyl semimetal in T\textsubscript{d}-MoTe\textsubscript{2}}},\ }\href@noop {} {\bibfield  {journal} {\bibinfo  {journal} {Physical Review Letters}\ }\textbf {\bibinfo {volume} {121}},\ \bibinfo {pages} {136401} (\bibinfo {year}
  {2018})}\BibitemShut {NoStop}%
\bibitem [{\citenamefont {Singh}\ \emph {et~al.}(2020)\citenamefont {Singh}, \citenamefont {Kim}, \citenamefont {Rabe},\ and\ \citenamefont {Vanderbilt}}]{singh2020engineering}%
  \BibitemOpen
  \bibfield  {author} {\bibinfo {author} {\bibfnamefont {S.}~\bibnamefont {Singh}}, \bibinfo {author} {\bibfnamefont {J.}~\bibnamefont {Kim}}, \bibinfo {author} {\bibfnamefont {K.~M.}\ \bibnamefont {Rabe}},\ and\ \bibinfo {author} {\bibfnamefont {D.}~\bibnamefont {Vanderbilt}},\ }\bibfield  {title} {\bibinfo {title} {{Engineering Weyl phases and nonlinear Hall effects in T\textsubscript{d}-MoTe\textsubscript{2}}},\ }\href@noop {} {\bibfield  {journal} {\bibinfo  {journal} {Physical Review Letters}\ }\textbf {\bibinfo {volume} {125}},\ \bibinfo {pages} {046402} (\bibinfo {year} {2020})}\BibitemShut {NoStop}%
\bibitem [{\citenamefont {Lee}\ \emph {et~al.}(2026)\citenamefont {Lee}, \citenamefont {Kang}, \citenamefont {Kim}, \citenamefont {Cho}, \citenamefont {Kim}, \citenamefont {Cheon},\ and\ \citenamefont {Park}}]{lee2026symmetry}%
  \BibitemOpen
  \bibfield  {author} {\bibinfo {author} {\bibfnamefont {S.}~\bibnamefont {Lee}}, \bibinfo {author} {\bibfnamefont {M.}~\bibnamefont {Kang}}, \bibinfo {author} {\bibfnamefont {J.}~\bibnamefont {Kim}}, \bibinfo {author} {\bibfnamefont {S.}~\bibnamefont {Cho}}, \bibinfo {author} {\bibfnamefont {D.~Y.}\ \bibnamefont {Kim}}, \bibinfo {author} {\bibfnamefont {S.}~\bibnamefont {Cheon}},\ and\ \bibinfo {author} {\bibfnamefont {T.}~\bibnamefont {Park}},\ }\bibfield  {title} {\bibinfo {title} {{Symmetry-controlled multi-gap superconductivity and higher-order topological phases of MoTe\textsubscript{2}}},\ }\href@noop {} {\bibfield  {journal} {\bibinfo  {journal} {Nature Communications}\ }\textbf {\bibinfo {volume} {17}},\ \bibinfo {pages} {7685} (\bibinfo {year} {2026})}\BibitemShut {NoStop}%
\bibitem [{\citenamefont {Naidyuk}\ \emph {et~al.}(2018)\citenamefont {Naidyuk}, \citenamefont {Kvitnitskaya}, \citenamefont {Bashlakov}, \citenamefont {Aswartham}, \citenamefont {Morozov}, \citenamefont {Chernyavskii}, \citenamefont {Fuchs}, \citenamefont {Drechsler}, \citenamefont {H{\"u}hne}, \citenamefont {Nielsch} \emph {et~al.}}]{naidyuk2018surface}%
  \BibitemOpen
  \bibfield  {author} {\bibinfo {author} {\bibfnamefont {Y.}~\bibnamefont {Naidyuk}}, \bibinfo {author} {\bibfnamefont {O.}~\bibnamefont {Kvitnitskaya}}, \bibinfo {author} {\bibfnamefont {D.}~\bibnamefont {Bashlakov}}, \bibinfo {author} {\bibfnamefont {S.}~\bibnamefont {Aswartham}}, \bibinfo {author} {\bibfnamefont {I.}~\bibnamefont {Morozov}}, \bibinfo {author} {\bibfnamefont {I.}~\bibnamefont {Chernyavskii}}, \bibinfo {author} {\bibfnamefont {G.}~\bibnamefont {Fuchs}}, \bibinfo {author} {\bibfnamefont {S.-L.}\ \bibnamefont {Drechsler}}, \bibinfo {author} {\bibfnamefont {R.}~\bibnamefont {H{\"u}hne}}, \bibinfo {author} {\bibfnamefont {K.}~\bibnamefont {Nielsch}}, \emph {et~al.},\ }\bibfield  {title} {\bibinfo {title} {{Surface superconductivity in the Weyl semimetal MoTe\textsubscript{2} detected by point contact spectroscopy}},\ }\href@noop {} {\bibfield  {journal} {\bibinfo  {journal} {2D Materials}\ }\textbf {\bibinfo {volume} {5}},\ \bibinfo {pages} {045014} (\bibinfo {year} {2018})}\BibitemShut
  {NoStop}%
\bibitem [{\citenamefont {Mu}\ \emph {et~al.}(2021)\citenamefont {Mu}, \citenamefont {Fan}, \citenamefont {Borrmann}, \citenamefont {Schnelle}, \citenamefont {Sun}, \citenamefont {Felser},\ and\ \citenamefont {Medvedev}}]{mu2021pressure}%
  \BibitemOpen
  \bibfield  {author} {\bibinfo {author} {\bibfnamefont {Q.-G.}\ \bibnamefont {Mu}}, \bibinfo {author} {\bibfnamefont {F.-R.}\ \bibnamefont {Fan}}, \bibinfo {author} {\bibfnamefont {H.}~\bibnamefont {Borrmann}}, \bibinfo {author} {\bibfnamefont {W.}~\bibnamefont {Schnelle}}, \bibinfo {author} {\bibfnamefont {Y.}~\bibnamefont {Sun}}, \bibinfo {author} {\bibfnamefont {C.}~\bibnamefont {Felser}},\ and\ \bibinfo {author} {\bibfnamefont {S.}~\bibnamefont {Medvedev}},\ }\bibfield  {title} {\bibinfo {title} {{Pressure-induced superconductivity and modification of Fermi surface in type-II Weyl semimetal NbIrTe\textsubscript{4}}},\ }\href@noop {} {\bibfield  {journal} {\bibinfo  {journal} {npj Quantum Materials}\ }\textbf {\bibinfo {volume} {6}},\ \bibinfo {pages} {55} (\bibinfo {year} {2021})}\BibitemShut {NoStop}%
\bibitem [{\citenamefont {Zhang}\ \emph {et~al.}(2024)\citenamefont {Zhang}, \citenamefont {Xu}, \citenamefont {Le}, \citenamefont {Chen}, \citenamefont {Ye}, \citenamefont {Shi}, \citenamefont {Luo}, \citenamefont {Shi},\ and\ \citenamefont {Lu}}]{zhang2024multigap}%
  \BibitemOpen
  \bibfield  {author} {\bibinfo {author} {\bibfnamefont {D.}~\bibnamefont {Zhang}}, \bibinfo {author} {\bibfnamefont {Z.}~\bibnamefont {Xu}}, \bibinfo {author} {\bibfnamefont {T.}~\bibnamefont {Le}}, \bibinfo {author} {\bibfnamefont {C.}~\bibnamefont {Chen}}, \bibinfo {author} {\bibfnamefont {G.}~\bibnamefont {Ye}}, \bibinfo {author} {\bibfnamefont {F.}~\bibnamefont {Shi}}, \bibinfo {author} {\bibfnamefont {S.}~\bibnamefont {Luo}}, \bibinfo {author} {\bibfnamefont {Y.}~\bibnamefont {Shi}},\ and\ \bibinfo {author} {\bibfnamefont {X.}~\bibnamefont {Lu}},\ }\bibfield  {title} {\bibinfo {title} {{Multigap s-wave superconductivity emerging in the 1T\textsuperscript{'} phase of MoTe\textsubscript{2} under hydrostatic pressure}},\ }\href@noop {} {\bibfield  {journal} {\bibinfo  {journal} {Physical Review B}\ }\textbf {\bibinfo {volume} {109}},\ \bibinfo {pages} {144506} (\bibinfo {year} {2024})}\BibitemShut {NoStop}%
\bibitem [{\citenamefont {Guguchia}\ \emph {et~al.}(2017)\citenamefont {Guguchia}, \citenamefont {von Rohr}, \citenamefont {Shermadini}, \citenamefont {Lee}, \citenamefont {Banerjee}, \citenamefont {Wieteska}, \citenamefont {Marianetti}, \citenamefont {Frandsen}, \citenamefont {Luetkens}, \citenamefont {Gong} \emph {et~al.}}]{guguchia2017signatures}%
  \BibitemOpen
  \bibfield  {author} {\bibinfo {author} {\bibfnamefont {Z.}~\bibnamefont {Guguchia}}, \bibinfo {author} {\bibfnamefont {F.}~\bibnamefont {von Rohr}}, \bibinfo {author} {\bibfnamefont {Z.}~\bibnamefont {Shermadini}}, \bibinfo {author} {\bibfnamefont {A.~T.}\ \bibnamefont {Lee}}, \bibinfo {author} {\bibfnamefont {S.}~\bibnamefont {Banerjee}}, \bibinfo {author} {\bibfnamefont {A.}~\bibnamefont {Wieteska}}, \bibinfo {author} {\bibfnamefont {C.}~\bibnamefont {Marianetti}}, \bibinfo {author} {\bibfnamefont {B.}~\bibnamefont {Frandsen}}, \bibinfo {author} {\bibfnamefont {H.}~\bibnamefont {Luetkens}}, \bibinfo {author} {\bibfnamefont {Z.}~\bibnamefont {Gong}}, \emph {et~al.},\ }\bibfield  {title} {\bibinfo {title} {{Signatures of the topological s\textsuperscript{+-} superconducting order parameter in the type-II Weyl semimetal T\textsubscript{d}-MoTe\textsubscript{2}}},\ }\href@noop {} {\bibfield  {journal} {\bibinfo  {journal} {Nature Communications}\ }\textbf {\bibinfo {volume} {8}},\ \bibinfo {pages} {1082}
  (\bibinfo {year} {2017})}\BibitemShut {NoStop}%
\bibitem [{\citenamefont {Luo}\ \emph {et~al.}(2020)\citenamefont {Luo}, \citenamefont {Li}, \citenamefont {Zhang}, \citenamefont {Ji}, \citenamefont {Wang}, \citenamefont {Shan}, \citenamefont {Zhang}, \citenamefont {Cai}, \citenamefont {Liu}, \citenamefont {Wang} \emph {et~al.}}]{luo2020possible}%
  \BibitemOpen
  \bibfield  {author} {\bibinfo {author} {\bibfnamefont {J.}~\bibnamefont {Luo}}, \bibinfo {author} {\bibfnamefont {Y.}~\bibnamefont {Li}}, \bibinfo {author} {\bibfnamefont {J.}~\bibnamefont {Zhang}}, \bibinfo {author} {\bibfnamefont {H.}~\bibnamefont {Ji}}, \bibinfo {author} {\bibfnamefont {H.}~\bibnamefont {Wang}}, \bibinfo {author} {\bibfnamefont {J.-Y.}\ \bibnamefont {Shan}}, \bibinfo {author} {\bibfnamefont {C.}~\bibnamefont {Zhang}}, \bibinfo {author} {\bibfnamefont {C.}~\bibnamefont {Cai}}, \bibinfo {author} {\bibfnamefont {J.}~\bibnamefont {Liu}}, \bibinfo {author} {\bibfnamefont {Y.}~\bibnamefont {Wang}}, \emph {et~al.},\ }\bibfield  {title} {\bibinfo {title} {{Possible unconventional two-band superconductivity in MoTe\textsubscript{2}}},\ }\href@noop {} {\bibfield  {journal} {\bibinfo  {journal} {Physical Review B}\ }\textbf {\bibinfo {volume} {102}},\ \bibinfo {pages} {064502} (\bibinfo {year} {2020})}\BibitemShut {NoStop}%
\bibitem [{\citenamefont {Kuibarov}\ \emph {et~al.}(2024)\citenamefont {Kuibarov}, \citenamefont {Suvorov}, \citenamefont {Vocaturo}, \citenamefont {Fedorov}, \citenamefont {Lou}, \citenamefont {Merkwitz}, \citenamefont {Voroshnin}, \citenamefont {Facio}, \citenamefont {Koepernik}, \citenamefont {Yaresko} \emph {et~al.}}]{kuibarov2024evidence}%
  \BibitemOpen
  \bibfield  {author} {\bibinfo {author} {\bibfnamefont {A.}~\bibnamefont {Kuibarov}}, \bibinfo {author} {\bibfnamefont {O.}~\bibnamefont {Suvorov}}, \bibinfo {author} {\bibfnamefont {R.}~\bibnamefont {Vocaturo}}, \bibinfo {author} {\bibfnamefont {A.}~\bibnamefont {Fedorov}}, \bibinfo {author} {\bibfnamefont {R.}~\bibnamefont {Lou}}, \bibinfo {author} {\bibfnamefont {L.}~\bibnamefont {Merkwitz}}, \bibinfo {author} {\bibfnamefont {V.}~\bibnamefont {Voroshnin}}, \bibinfo {author} {\bibfnamefont {J.~I.}\ \bibnamefont {Facio}}, \bibinfo {author} {\bibfnamefont {K.}~\bibnamefont {Koepernik}}, \bibinfo {author} {\bibfnamefont {A.}~\bibnamefont {Yaresko}}, \emph {et~al.},\ }\bibfield  {title} {\bibinfo {title} {{Evidence of superconducting Fermi arcs}},\ }\href@noop {} {\bibfield  {journal} {\bibinfo  {journal} {Nature}\ }\textbf {\bibinfo {volume} {626}},\ \bibinfo {pages} {294} (\bibinfo {year} {2024})}\BibitemShut {NoStop}%
\bibitem [{\citenamefont {Zhang}\ \emph {et~al.}(2025)\citenamefont {Zhang}, \citenamefont {Chen}, \citenamefont {Huang}, \citenamefont {Wang}, \citenamefont {Han}, \citenamefont {Ma}, \citenamefont {Zhu}, \citenamefont {Ning}, \citenamefont {Shen}, \citenamefont {Huan} \emph {et~al.}}]{zhang2025atomic}%
  \BibitemOpen
  \bibfield  {author} {\bibinfo {author} {\bibfnamefont {H.}~\bibnamefont {Zhang}}, \bibinfo {author} {\bibfnamefont {H.}~\bibnamefont {Chen}}, \bibinfo {author} {\bibfnamefont {Z.}~\bibnamefont {Huang}}, \bibinfo {author} {\bibfnamefont {Z.-A.}\ \bibnamefont {Wang}}, \bibinfo {author} {\bibfnamefont {G.}~\bibnamefont {Han}}, \bibinfo {author} {\bibfnamefont {R.}~\bibnamefont {Ma}}, \bibinfo {author} {\bibfnamefont {X.}~\bibnamefont {Zhu}}, \bibinfo {author} {\bibfnamefont {W.}~\bibnamefont {Ning}}, \bibinfo {author} {\bibfnamefont {C.}~\bibnamefont {Shen}}, \bibinfo {author} {\bibfnamefont {Q.}~\bibnamefont {Huan}}, \emph {et~al.},\ }\bibfield  {title} {\bibinfo {title} {{Atomic Visualization of Bulk and Surface Superconductivity in Weyl Semimetal $\gamma$-PtBi\textsubscript{2}}},\ }\href@noop {} {\bibfield  {journal} {\bibinfo  {journal} {Chinese Physics Letters}\ }\textbf {\bibinfo {volume} {42}},\ \bibinfo {pages} {120708} (\bibinfo {year} {2025})}\BibitemShut {NoStop}%
\bibitem [{\citenamefont {Schimmel}\ \emph {et~al.}(2024)\citenamefont {Schimmel}, \citenamefont {Fasano}, \citenamefont {Hoffmann}, \citenamefont {Besproswanny}, \citenamefont {Corredor~Bohorquez}, \citenamefont {Puig}, \citenamefont {Elshalem}, \citenamefont {Kalisky}, \citenamefont {Shipunov}, \citenamefont {Baumann} \emph {et~al.}}]{schimmel2024surface}%
  \BibitemOpen
  \bibfield  {author} {\bibinfo {author} {\bibfnamefont {S.}~\bibnamefont {Schimmel}}, \bibinfo {author} {\bibfnamefont {Y.}~\bibnamefont {Fasano}}, \bibinfo {author} {\bibfnamefont {S.}~\bibnamefont {Hoffmann}}, \bibinfo {author} {\bibfnamefont {J.}~\bibnamefont {Besproswanny}}, \bibinfo {author} {\bibfnamefont {L.~T.}\ \bibnamefont {Corredor~Bohorquez}}, \bibinfo {author} {\bibfnamefont {J.}~\bibnamefont {Puig}}, \bibinfo {author} {\bibfnamefont {B.-C.}\ \bibnamefont {Elshalem}}, \bibinfo {author} {\bibfnamefont {B.}~\bibnamefont {Kalisky}}, \bibinfo {author} {\bibfnamefont {G.}~\bibnamefont {Shipunov}}, \bibinfo {author} {\bibfnamefont {D.}~\bibnamefont {Baumann}}, \emph {et~al.},\ }\bibfield  {title} {\bibinfo {title} {{Surface superconductivity in the topological Weyl semimetal $t$-PtBi\textsubscript{2}}},\ }\href@noop {} {\bibfield  {journal} {\bibinfo  {journal} {Nature Communications}\ }\textbf {\bibinfo {volume} {15}},\ \bibinfo {pages} {9895} (\bibinfo {year} {2024})}\BibitemShut {NoStop}%
\bibitem [{\citenamefont {Belopolski}\ \emph {et~al.}(2017)\citenamefont {Belopolski}, \citenamefont {Yu}, \citenamefont {Sanchez}, \citenamefont {Ishida}, \citenamefont {Chang}, \citenamefont {Zhang}, \citenamefont {Xu}, \citenamefont {Zheng}, \citenamefont {Chang}, \citenamefont {Bian} \emph {et~al.}}]{belopolski2017signatures}%
  \BibitemOpen
  \bibfield  {author} {\bibinfo {author} {\bibfnamefont {I.}~\bibnamefont {Belopolski}}, \bibinfo {author} {\bibfnamefont {P.}~\bibnamefont {Yu}}, \bibinfo {author} {\bibfnamefont {D.~S.}\ \bibnamefont {Sanchez}}, \bibinfo {author} {\bibfnamefont {Y.}~\bibnamefont {Ishida}}, \bibinfo {author} {\bibfnamefont {T.-R.}\ \bibnamefont {Chang}}, \bibinfo {author} {\bibfnamefont {S.~S.}\ \bibnamefont {Zhang}}, \bibinfo {author} {\bibfnamefont {S.-Y.}\ \bibnamefont {Xu}}, \bibinfo {author} {\bibfnamefont {H.}~\bibnamefont {Zheng}}, \bibinfo {author} {\bibfnamefont {G.}~\bibnamefont {Chang}}, \bibinfo {author} {\bibfnamefont {G.}~\bibnamefont {Bian}}, \emph {et~al.},\ }\bibfield  {title} {\bibinfo {title} {{Signatures of a time-reversal symmetric Weyl semimetal with only four Weyl points}},\ }\href@noop {} {\bibfield  {journal} {\bibinfo  {journal} {Nature Communications}\ }\textbf {\bibinfo {volume} {8}},\ \bibinfo {pages} {942} (\bibinfo {year} {2017})}\BibitemShut {NoStop}%
\bibitem [{\citenamefont {McCormick}\ \emph {et~al.}(2017)\citenamefont {McCormick}, \citenamefont {Kimchi},\ and\ \citenamefont {Trivedi}}]{mccormick2017minimal}%
  \BibitemOpen
  \bibfield  {author} {\bibinfo {author} {\bibfnamefont {T.~M.}\ \bibnamefont {McCormick}}, \bibinfo {author} {\bibfnamefont {I.}~\bibnamefont {Kimchi}},\ and\ \bibinfo {author} {\bibfnamefont {N.}~\bibnamefont {Trivedi}},\ }\bibfield  {title} {\bibinfo {title} {{Minimal models for topological Weyl semimetals}},\ }\href@noop {} {\bibfield  {journal} {\bibinfo  {journal} {Physical Review B}\ }\textbf {\bibinfo {volume} {95}},\ \bibinfo {pages} {075133} (\bibinfo {year} {2017})}\BibitemShut {NoStop}%
  \bibitem [{\citenamefont {Murakami}(2007)}]{murakami2007phase}%
  \BibitemOpen
  \bibfield  {author} {\bibinfo {author} {\bibfnamefont {S.}~\bibnamefont {Murakami}},\ }\bibfield  {title} {\bibinfo {title} {{Phase transition between the quantum spin Hall and insulator phases in 3D: emergence of a topological gapless phase}},\ }\href@noop {} {\bibfield  {journal} {\bibinfo  {journal} {New Journal of Physics}\ }\textbf {\bibinfo {volume} {9}},\ \bibinfo {pages} {356} (\bibinfo {year} {2007})}\BibitemShut {NoStop}%
\bibitem [{sup()}]{supple}%
  \BibitemOpen
  \href@noop {} {}\bibinfo {note} {See Supplemental Material [URL] for detailed derivation and calculations.}\BibitemShut {Stop}%
\bibitem [{\citenamefont {Schnyder}\ \emph {et~al.}(2008)\citenamefont {Schnyder}, \citenamefont {Ryu}, \citenamefont {Furusaki},\ and\ \citenamefont {Ludwig}}]{schnyder2008classification}%
  \BibitemOpen
  \bibfield  {author} {\bibinfo {author} {\bibfnamefont {A.~P.}\ \bibnamefont {Schnyder}}, \bibinfo {author} {\bibfnamefont {S.}~\bibnamefont {Ryu}}, \bibinfo {author} {\bibfnamefont {A.}~\bibnamefont {Furusaki}},\ and\ \bibinfo {author} {\bibfnamefont {A.~W.}\ \bibnamefont {Ludwig}},\ }\bibfield  {title} {\bibinfo {title} {{Classification of topological insulators and superconductors in three spatial dimensions}},\ }\href@noop {} {\bibfield  {journal} {\bibinfo  {journal} {Physical Review B}\ }\textbf {\bibinfo {volume} {78}},\ \bibinfo {pages} {195125} (\bibinfo {year} {2008})}\BibitemShut {NoStop}%
\bibitem [{\citenamefont {Chiu}\ \emph {et~al.}(2016)\citenamefont {Chiu}, \citenamefont {Teo}, \citenamefont {Schnyder},\ and\ \citenamefont {Ryu}}]{chiu2016classification}%
  \BibitemOpen
  \bibfield  {author} {\bibinfo {author} {\bibfnamefont {C.-K.}\ \bibnamefont {Chiu}}, \bibinfo {author} {\bibfnamefont {J.~C.}\ \bibnamefont {Teo}}, \bibinfo {author} {\bibfnamefont {A.~P.}\ \bibnamefont {Schnyder}},\ and\ \bibinfo {author} {\bibfnamefont {S.}~\bibnamefont {Ryu}},\ }\bibfield  {title} {\bibinfo {title} {{Classification of topological quantum matter with symmetries}},\ }\href@noop {} {\bibfield  {journal} {\bibinfo  {journal} {Reviews of Modern Physics}\ }\textbf {\bibinfo {volume} {88}},\ \bibinfo {pages} {035005} (\bibinfo {year} {2016})}\BibitemShut {NoStop}%
\bibitem [{\citenamefont {Sigrist}\ and\ \citenamefont {Ueda}(1991)}]{sigrist1991phenomenological}%
  \BibitemOpen
  \bibfield  {author} {\bibinfo {author} {\bibfnamefont {M.}~\bibnamefont {Sigrist}}\ and\ \bibinfo {author} {\bibfnamefont {K.}~\bibnamefont {Ueda}},\ }\bibfield  {title} {\bibinfo {title} {{Phenomenological theory of unconventional superconductivity}},\ }\href@noop {} {\bibfield  {journal} {\bibinfo  {journal} {Reviews of Modern Physics}\ }\textbf {\bibinfo {volume} {63}},\ \bibinfo {pages} {239} (\bibinfo {year} {1991})}\BibitemShut {NoStop}%
\bibitem [{\citenamefont {Lee}\ \emph {et~al.}(2018)\citenamefont {Lee}, \citenamefont {Jang}, \citenamefont {Kim}, \citenamefont {Jung}, \citenamefont {Kim}, \citenamefont {Cho}, \citenamefont {Kim}, \citenamefont {Rhee}, \citenamefont {Park},\ and\ \citenamefont {Park}}]{lee2018origin}%
  \BibitemOpen
  \bibfield  {author} {\bibinfo {author} {\bibfnamefont {S.}~\bibnamefont {Lee}}, \bibinfo {author} {\bibfnamefont {J.}~\bibnamefont {Jang}}, \bibinfo {author} {\bibfnamefont {S.-I.}\ \bibnamefont {Kim}}, \bibinfo {author} {\bibfnamefont {S.-G.}\ \bibnamefont {Jung}}, \bibinfo {author} {\bibfnamefont {J.}~\bibnamefont {Kim}}, \bibinfo {author} {\bibfnamefont {S.}~\bibnamefont {Cho}}, \bibinfo {author} {\bibfnamefont {S.~W.}\ \bibnamefont {Kim}}, \bibinfo {author} {\bibfnamefont {J.~Y.}\ \bibnamefont {Rhee}}, \bibinfo {author} {\bibfnamefont {K.-S.}\ \bibnamefont {Park}},\ and\ \bibinfo {author} {\bibfnamefont {T.}~\bibnamefont {Park}},\ }\bibfield  {title} {\bibinfo {title} {{Origin of extremely large magnetoresistance in the candidate type-II Weyl semimetal MoTe\textsubscript{2- x}}},\ }\href@noop {} {\bibfield  {journal} {\bibinfo  {journal} {Scientific Reports}\ }\textbf {\bibinfo {volume} {8}},\ \bibinfo {pages} {13937} (\bibinfo {year} {2018})}\BibitemShut {NoStop}%
\bibitem [{\citenamefont {Szab{\'o}}\ \emph {et~al.}(2001)\citenamefont {Szab{\'o}}, \citenamefont {Samuely}, \citenamefont {Ka{\v{c}}mar{\v{c}}{\'\i}k}, \citenamefont {Klein}, \citenamefont {Marcus}, \citenamefont {Fruchart}, \citenamefont {Miraglia}, \citenamefont {Marcenat},\ and\ \citenamefont {Jansen}}]{szabo2001evidence}%
  \BibitemOpen
  \bibfield  {author} {\bibinfo {author} {\bibfnamefont {P.}~\bibnamefont {Szab{\'o}}}, \bibinfo {author} {\bibfnamefont {P.}~\bibnamefont {Samuely}}, \bibinfo {author} {\bibfnamefont {J.}~\bibnamefont {Ka{\v{c}}mar{\v{c}}{\'\i}k}}, \bibinfo {author} {\bibfnamefont {T.}~\bibnamefont {Klein}}, \bibinfo {author} {\bibfnamefont {J.}~\bibnamefont {Marcus}}, \bibinfo {author} {\bibfnamefont {D.}~\bibnamefont {Fruchart}}, \bibinfo {author} {\bibfnamefont {S.}~\bibnamefont {Miraglia}}, \bibinfo {author} {\bibfnamefont {C.}~\bibnamefont {Marcenat}},\ and\ \bibinfo {author} {\bibfnamefont {A.~G.~M.}\ \bibnamefont {Jansen}},\ }\bibfield  {title} {\bibinfo {title} {{Evidence for Two Superconducting Energy Gaps in MgB\textsubscript{2} by Point-Contact Spectroscopy}},\ }\href@noop {} {\bibfield  {journal} {\bibinfo  {journal} {Physical Review Letters}\ }\textbf {\bibinfo {volume} {87}},\ \bibinfo {pages} {137005} (\bibinfo {year} {2001})}\BibitemShut {NoStop}%
\bibitem [{\citenamefont {Bouquet}\ \emph {et~al.}(2001)\citenamefont {Bouquet}, \citenamefont {Fisher}, \citenamefont {Phillips}, \citenamefont {Hinks},\ and\ \citenamefont {Jorgensen}}]{bouquet2001specific}%
  \BibitemOpen
  \bibfield  {author} {\bibinfo {author} {\bibfnamefont {F.}~\bibnamefont {Bouquet}}, \bibinfo {author} {\bibfnamefont {R.~A.}\ \bibnamefont {Fisher}}, \bibinfo {author} {\bibfnamefont {N.~E.}\ \bibnamefont {Phillips}}, \bibinfo {author} {\bibfnamefont {D.~G.}\ \bibnamefont {Hinks}},\ and\ \bibinfo {author} {\bibfnamefont {J.~D.}\ \bibnamefont {Jorgensen}},\ }\bibfield  {title} {\bibinfo {title} {{Specific Heat of Mg\textsuperscript{11}B\textsubscript{2}: Evidence for a Second Energy Gap}},\ }\href@noop {} {\bibfield  {journal} {\bibinfo  {journal} {Physical Review Letters}\ }\textbf {\bibinfo {volume} {87}},\ \bibinfo {pages} {047001} (\bibinfo {year} {2001})}\BibitemShut {NoStop}%
\bibitem [{\citenamefont {Huang}\ \emph {et~al.}(2007)\citenamefont {Huang}, \citenamefont {Lin}, \citenamefont {Chang}, \citenamefont {Sun}, \citenamefont {Shen}, \citenamefont {Chou}, \citenamefont {Berger}, \citenamefont {Lee},\ and\ \citenamefont {Yang}}]{huang2007experimental}%
  \BibitemOpen
  \bibfield  {author} {\bibinfo {author} {\bibfnamefont {C.}~\bibnamefont {Huang}}, \bibinfo {author} {\bibfnamefont {J.-Y.}\ \bibnamefont {Lin}}, \bibinfo {author} {\bibfnamefont {Y.}~\bibnamefont {Chang}}, \bibinfo {author} {\bibfnamefont {C.}~\bibnamefont {Sun}}, \bibinfo {author} {\bibfnamefont {H.}~\bibnamefont {Shen}}, \bibinfo {author} {\bibfnamefont {C.}~\bibnamefont {Chou}}, \bibinfo {author} {\bibfnamefont {H.}~\bibnamefont {Berger}}, \bibinfo {author} {\bibfnamefont {T.}~\bibnamefont {Lee}},\ and\ \bibinfo {author} {\bibfnamefont {H.}~\bibnamefont {Yang}},\ }\bibfield  {title} {\bibinfo {title} {{Experimental evidence for a two-gap structure of superconducting NbSe\textsubscript{2}: A specific-heat study in external magnetic fields}},\ }\href@noop {} {\bibfield  {journal} {\bibinfo  {journal} {Physical Review B}\ }\textbf {\bibinfo {volume} {76}},\ \bibinfo {pages} {212504} (\bibinfo {year} {2007})}\BibitemShut {NoStop}%
\bibitem [{\citenamefont {Boaknin}\ \emph {et~al.}(2003)\citenamefont {Boaknin}, \citenamefont {Tanatar}, \citenamefont {Paglione}, \citenamefont {Hawthorn}, \citenamefont {Ronning}, \citenamefont {Hill}, \citenamefont {Sutherland}, \citenamefont {Taillefer}, \citenamefont {Sonier}, \citenamefont {Hayden} \emph {et~al.}}]{boaknin2003heat}%
  \BibitemOpen
  \bibfield  {author} {\bibinfo {author} {\bibfnamefont {E.}~\bibnamefont {Boaknin}}, \bibinfo {author} {\bibfnamefont {M.}~\bibnamefont {Tanatar}}, \bibinfo {author} {\bibfnamefont {J.}~\bibnamefont {Paglione}}, \bibinfo {author} {\bibfnamefont {D.}~\bibnamefont {Hawthorn}}, \bibinfo {author} {\bibfnamefont {F.}~\bibnamefont {Ronning}}, \bibinfo {author} {\bibfnamefont {R.}~\bibnamefont {Hill}}, \bibinfo {author} {\bibfnamefont {M.}~\bibnamefont {Sutherland}}, \bibinfo {author} {\bibfnamefont {L.}~\bibnamefont {Taillefer}}, \bibinfo {author} {\bibfnamefont {J.}~\bibnamefont {Sonier}}, \bibinfo {author} {\bibfnamefont {S.}~\bibnamefont {Hayden}}, \emph {et~al.},\ }\bibfield  {title} {\bibinfo {title} {{Heat conduction in the vortex state of NbSe\textsubscript{2}: Evidence for multiband superconductivity}},\ }\href@noop {} {\bibfield  {journal} {\bibinfo  {journal} {Physical Review Letters}\ }\textbf {\bibinfo {volume} {90}},\ \bibinfo {pages} {117003} (\bibinfo {year} {2003})}\BibitemShut {NoStop}%
\bibitem [{\citenamefont {Ka{\v{c}}mar{\v{c}}{\'\i}k}\ \emph {et~al.}(2010)\citenamefont {Ka{\v{c}}mar{\v{c}}{\'\i}k}, \citenamefont {Pribulov{\'a}}, \citenamefont {Marcenat}, \citenamefont {Klein}, \citenamefont {Rodi{\`e}re}, \citenamefont {Cario},\ and\ \citenamefont {Samuely}}]{kavcmarvcik2010studies}%
  \BibitemOpen
  \bibfield  {author} {\bibinfo {author} {\bibfnamefont {J.}~\bibnamefont {Ka{\v{c}}mar{\v{c}}{\'\i}k}}, \bibinfo {author} {\bibfnamefont {Z.}~\bibnamefont {Pribulov{\'a}}}, \bibinfo {author} {\bibfnamefont {C.}~\bibnamefont {Marcenat}}, \bibinfo {author} {\bibfnamefont {T.}~\bibnamefont {Klein}}, \bibinfo {author} {\bibfnamefont {P.}~\bibnamefont {Rodi{\`e}re}}, \bibinfo {author} {\bibfnamefont {L.}~\bibnamefont {Cario}},\ and\ \bibinfo {author} {\bibfnamefont {P.}~\bibnamefont {Samuely}},\ }\bibfield  {title} {\bibinfo {title} {{Studies on two-gap superconductivity in 2H--NbS\textsubscript{2}}},\ }\href@noop {} {\bibfield  {journal} {\bibinfo  {journal} {Physica C: Superconductivity and its Applications}\ }\textbf {\bibinfo {volume} {470}},\ \bibinfo {pages} {S719} (\bibinfo {year} {2010})}\BibitemShut {NoStop}%
\bibitem [{\citenamefont {Perali}\ \emph {et~al.}(1996)\citenamefont {Perali}, \citenamefont {Bianconi}, \citenamefont {Lanzara},\ and\ \citenamefont {Saini}}]{perali1996gap}%
  \BibitemOpen
  \bibfield  {author} {\bibinfo {author} {\bibfnamefont {A.}~\bibnamefont {Perali}}, \bibinfo {author} {\bibfnamefont {A.}~\bibnamefont {Bianconi}}, \bibinfo {author} {\bibfnamefont {A.}~\bibnamefont {Lanzara}},\ and\ \bibinfo {author} {\bibfnamefont {N.~L.}\ \bibnamefont {Saini}},\ }\bibfield  {title} {\bibinfo {title} {{The gap amplification at a shape resonance in a superlattice of quantum stripes: A mechanism for high T\textsubscript{c}}},\ }\href@noop {} {\bibfield  {journal} {\bibinfo  {journal} {Solid State Communications}\ }\textbf {\bibinfo {volume} {100}},\ \bibinfo {pages} {181} (\bibinfo {year} {1996})}\BibitemShut {NoStop}%
\bibitem [{\citenamefont {Innocenti}\ \emph {et~al.}(2010)\citenamefont {Innocenti}, \citenamefont {Poccia}, \citenamefont {Ricci}, \citenamefont {Valletta}, \citenamefont {Caprara}, \citenamefont {Perali},\ and\ \citenamefont {Bianconi}}]{innocenti2010resonant}%
  \BibitemOpen
  \bibfield  {author} {\bibinfo {author} {\bibfnamefont {D.}~\bibnamefont {Innocenti}}, \bibinfo {author} {\bibfnamefont {N.}~\bibnamefont {Poccia}}, \bibinfo {author} {\bibfnamefont {A.}~\bibnamefont {Ricci}}, \bibinfo {author} {\bibfnamefont {A.}~\bibnamefont {Valletta}}, \bibinfo {author} {\bibfnamefont {S.}~\bibnamefont {Caprara}}, \bibinfo {author} {\bibfnamefont {A.}~\bibnamefont {Perali}},\ and\ \bibinfo {author} {\bibfnamefont {A.}~\bibnamefont {Bianconi}},\ }\bibfield  {title} {\bibinfo {title} {{Resonant and crossover phenomena in a multiband superconductor: Tuning the chemical potential near a band edge}},\ }\href@noop {} {\bibfield  {journal} {\bibinfo  {journal} {Physical Review B}\ }\textbf {\bibinfo {volume} {82}},\ \bibinfo {pages} {184528} (\bibinfo {year} {2010})}\BibitemShut {NoStop}%
\bibitem [{\citenamefont {Coldea}\ and\ \citenamefont {Watson}(2018)}]{coldea2018key}%
  \BibitemOpen
  \bibfield  {author} {\bibinfo {author} {\bibfnamefont {A.~I.}\ \bibnamefont {Coldea}}\ and\ \bibinfo {author} {\bibfnamefont {M.~D.}\ \bibnamefont {Watson}},\ }\bibfield  {title} {\bibinfo {title} {{The Key Ingredients of the Electronic Structure of FeSe}},\ }\href@noop {} {\bibfield  {journal} {\bibinfo  {journal} {Annual Review of Condensed Matter Physics}\ }\textbf {\bibinfo {volume} {9}},\ \bibinfo {pages} {125} (\bibinfo {year} {2018})}\BibitemShut {NoStop}%
\bibitem [{\citenamefont {Qi}\ \emph {et~al.}(2016)\citenamefont {Qi}, \citenamefont {Naumov}, \citenamefont {Ali}, \citenamefont {Rajamathi}, \citenamefont {Schnelle}, \citenamefont {Barkalov}, \citenamefont {Hanfland}, \citenamefont {Wu}, \citenamefont {Shekhar}, \citenamefont {Sun}, \citenamefont {S{\"u}{\ss}}, \citenamefont {Schmidt}, \citenamefont {Schwarz}, \citenamefont {Pippel}, \citenamefont {Werner}, \citenamefont {Hillebrand}, \citenamefont {F{\"o}rster}, \citenamefont {Kampert}, \citenamefont {Parkin}, \citenamefont {Cava}, \citenamefont {Felser}, \citenamefont {Yan},\ and\ \citenamefont {Medvedev}}]{qi2016superconductivity}%
  \BibitemOpen
  \bibfield  {author} {\bibinfo {author} {\bibfnamefont {Y.}~\bibnamefont {Qi}}, \bibinfo {author} {\bibfnamefont {P.~G.}\ \bibnamefont {Naumov}}, \bibinfo {author} {\bibfnamefont {M.~N.}\ \bibnamefont {Ali}}, \bibinfo {author} {\bibfnamefont {C.~R.}\ \bibnamefont {Rajamathi}}, \bibinfo {author} {\bibfnamefont {W.}~\bibnamefont {Schnelle}}, \bibinfo {author} {\bibfnamefont {O.}~\bibnamefont {Barkalov}}, \bibinfo {author} {\bibfnamefont {M.}~\bibnamefont {Hanfland}}, \bibinfo {author} {\bibfnamefont {S.-C.}\ \bibnamefont {Wu}}, \bibinfo {author} {\bibfnamefont {C.}~\bibnamefont {Shekhar}}, \bibinfo {author} {\bibfnamefont {Y.}~\bibnamefont {Sun}}, \bibinfo {author} {\bibfnamefont {V.}~\bibnamefont {S{\"u}{\ss}}}, \bibinfo {author} {\bibfnamefont {M.}~\bibnamefont {Schmidt}}, \bibinfo {author} {\bibfnamefont {U.}~\bibnamefont {Schwarz}}, \bibinfo {author} {\bibfnamefont {E.}~\bibnamefont {Pippel}}, \bibinfo {author} {\bibfnamefont {P.}~\bibnamefont {Werner}}, \bibinfo {author} {\bibfnamefont {R.}~\bibnamefont
  {Hillebrand}}, \bibinfo {author} {\bibfnamefont {T.}~\bibnamefont {F{\"o}rster}}, \bibinfo {author} {\bibfnamefont {E.}~\bibnamefont {Kampert}}, \bibinfo {author} {\bibfnamefont {S.}~\bibnamefont {Parkin}}, \bibinfo {author} {\bibfnamefont {R.~J.}\ \bibnamefont {Cava}}, \bibinfo {author} {\bibfnamefont {C.}~\bibnamefont {Felser}}, \bibinfo {author} {\bibfnamefont {B.}~\bibnamefont {Yan}},\ and\ \bibinfo {author} {\bibfnamefont {S.~A.}\ \bibnamefont {Medvedev}},\ }\bibfield  {title} {\bibinfo {title} {{Superconductivity in Weyl semimetal candidate MoTe\textsubscript{2}}},\ }\href@noop {} {\bibfield  {journal} {\bibinfo  {journal} {Nature Communications}\ }\textbf {\bibinfo {volume} {7}},\ \bibinfo {pages} {11038} (\bibinfo {year} {2016})}\BibitemShut {NoStop}%
\bibitem [{\citenamefont {Bahramy}\ \emph {et~al.}(2018)\citenamefont {Bahramy}, \citenamefont {Clark}, \citenamefont {Yang}, \citenamefont {Feng}, \citenamefont {Bawden}, \citenamefont {Riley}, \citenamefont {Markovi{\'c}}, \citenamefont {Mazzola}, \citenamefont {Sunko}, \citenamefont {Biswas} \emph {et~al.}}]{bahramy2018ubiquitous}%
  \BibitemOpen
  \bibfield  {author} {\bibinfo {author} {\bibfnamefont {M.}~\bibnamefont {Bahramy}}, \bibinfo {author} {\bibfnamefont {O.}~\bibnamefont {Clark}}, \bibinfo {author} {\bibfnamefont {B.-J.}\ \bibnamefont {Yang}}, \bibinfo {author} {\bibfnamefont {J.}~\bibnamefont {Feng}}, \bibinfo {author} {\bibfnamefont {L.}~\bibnamefont {Bawden}}, \bibinfo {author} {\bibfnamefont {J.}~\bibnamefont {Riley}}, \bibinfo {author} {\bibfnamefont {I.}~\bibnamefont {Markovi{\'c}}}, \bibinfo {author} {\bibfnamefont {F.}~\bibnamefont {Mazzola}}, \bibinfo {author} {\bibfnamefont {V.}~\bibnamefont {Sunko}}, \bibinfo {author} {\bibfnamefont {D.}~\bibnamefont {Biswas}}, \emph {et~al.},\ }\bibfield  {title} {\bibinfo {title} {{Ubiquitous formation of bulk Dirac cones and topological surface states from a single orbital manifold in transition-metal dichalcogenides}},\ }\href@noop {} {\bibfield  {journal} {\bibinfo  {journal} {Nature Materials}\ }\textbf {\bibinfo {volume} {17}},\ \bibinfo {pages} {21} (\bibinfo {year} {2018})}\BibitemShut
  {NoStop}%
\bibitem [{\citenamefont {Wang}\ \emph {et~al.}(2019)\citenamefont {Wang}, \citenamefont {Wieder}, \citenamefont {Li}, \citenamefont {Yan},\ and\ \citenamefont {Bernevig}}]{wang2019higher}%
  \BibitemOpen
  \bibfield  {author} {\bibinfo {author} {\bibfnamefont {Z.}~\bibnamefont {Wang}}, \bibinfo {author} {\bibfnamefont {B.~J.}\ \bibnamefont {Wieder}}, \bibinfo {author} {\bibfnamefont {J.}~\bibnamefont {Li}}, \bibinfo {author} {\bibfnamefont {B.}~\bibnamefont {Yan}},\ and\ \bibinfo {author} {\bibfnamefont {B.~A.}\ \bibnamefont {Bernevig}},\ }\bibfield  {title} {\bibinfo {title} {{Higher-Order Topology, Monopole Nodal Lines, and the Origin of Large Fermi Arcs in Transition Metal Dichalcogenides XTe\textsubscript{2} (X= Mo, W)}},\ }\href@noop {} {\bibfield  {journal} {\bibinfo  {journal} {Physical Review Letters}\ }\textbf {\bibinfo {volume} {123}},\ \bibinfo {pages} {186401} (\bibinfo {year} {2019})}\BibitemShut {NoStop}%
\bibitem [{\citenamefont {Hu}\ \emph {et~al.}(2020)\citenamefont {Hu}, \citenamefont {Yu}, \citenamefont {Lai}, \citenamefont {Sun}, \citenamefont {Balakirev}, \citenamefont {Zhang}, \citenamefont {Xie}, \citenamefont {Yip}, \citenamefont {Aulestia}, \citenamefont {Jha} \emph {et~al.}}]{hu2020detection}%
  \BibitemOpen
  \bibfield  {author} {\bibinfo {author} {\bibfnamefont {Y.}~\bibnamefont {Hu}}, \bibinfo {author} {\bibfnamefont {W.~C.}\ \bibnamefont {Yu}}, \bibinfo {author} {\bibfnamefont {K.~T.}\ \bibnamefont {Lai}}, \bibinfo {author} {\bibfnamefont {D.}~\bibnamefont {Sun}}, \bibinfo {author} {\bibfnamefont {F.~F.}\ \bibnamefont {Balakirev}}, \bibinfo {author} {\bibfnamefont {W.}~\bibnamefont {Zhang}}, \bibinfo {author} {\bibfnamefont {J.}~\bibnamefont {Xie}}, \bibinfo {author} {\bibfnamefont {K.}~\bibnamefont {Yip}}, \bibinfo {author} {\bibfnamefont {E.~P.}\ \bibnamefont {Aulestia}}, \bibinfo {author} {\bibfnamefont {R.}~\bibnamefont {Jha}}, \emph {et~al.},\ }\bibfield  {title} {\bibinfo {title} {{Detection of Hole Pockets in the Candidate Type-II Weyl Semimetal MoTe\textsubscript{2} from Shubnikov--de Haas Quantum Oscillations}},\ }\href@noop {} {\bibfield  {journal} {\bibinfo  {journal} {Physical Review Letters}\ }\textbf {\bibinfo {volume} {124}},\ \bibinfo {pages} {076402} (\bibinfo {year} {2020})}\BibitemShut
  {NoStop}%
\bibitem [{\citenamefont {Xu}\ \emph {et~al.}(2014)\citenamefont {Xu}, \citenamefont {Wang}, \citenamefont {Yan},\ and\ \citenamefont {Qi}}]{xu2014topological}%
  \BibitemOpen
  \bibfield  {author} {\bibinfo {author} {\bibfnamefont {G.}~\bibnamefont {Xu}}, \bibinfo {author} {\bibfnamefont {J.}~\bibnamefont {Wang}}, \bibinfo {author} {\bibfnamefont {B.}~\bibnamefont {Yan}},\ and\ \bibinfo {author} {\bibfnamefont {X.-L.}\ \bibnamefont {Qi}},\ }\bibfield  {title} {\bibinfo {title} {{Topological superconductivity at the edge of transition-metal dichalcogenides}},\ }\href@noop {} {\bibfield  {journal} {\bibinfo  {journal} {Physical Review B}\ }\textbf {\bibinfo {volume} {90}},\ \bibinfo {pages} {100505} (\bibinfo {year} {2014})}\BibitemShut {NoStop}%
\bibitem [{\citenamefont {Lane}\ and\ \citenamefont {Zhu}(2022)}]{lane2022identifying}%
  \BibitemOpen
  \bibfield  {author} {\bibinfo {author} {\bibfnamefont {C.}~\bibnamefont {Lane}}\ and\ \bibinfo {author} {\bibfnamefont {J.-X.}\ \bibnamefont {Zhu}},\ }\bibfield  {title} {\bibinfo {title} {{Identifying topological superconductivity in two-dimensional transition-metal dichalcogenides}},\ }\href@noop {} {\bibfield  {journal} {\bibinfo  {journal} {Physical Review Materials}\ }\textbf {\bibinfo {volume} {6}},\ \bibinfo {pages} {094001} (\bibinfo {year} {2022})}\BibitemShut {NoStop}%
\bibitem [{\citenamefont {Schrade}\ and\ \citenamefont {Fu}(2024)}]{schrade2024nematic}%
  \BibitemOpen
  \bibfield  {author} {\bibinfo {author} {\bibfnamefont {C.}~\bibnamefont {Schrade}}\ and\ \bibinfo {author} {\bibfnamefont {L.}~\bibnamefont {Fu}},\ }\bibfield  {title} {\bibinfo {title} {{Nematic, chiral, and topological superconductivity in twisted transition metal dichalcogenides}},\ }\href@noop {} {\bibfield  {journal} {\bibinfo  {journal} {Physical Review B}\ }\textbf {\bibinfo {volume} {110}},\ \bibinfo {pages} {035143} (\bibinfo {year} {2024})}\BibitemShut {NoStop}%
\end{thebibliography}
\end{document}